# Intergrain connectivity and resistive broadening in vortex state: a comparison between $MgB_2$, $NbSe_2$ and $Bi_2Sr_2Ca_2Cu_3O_{10}$ superconductors

S. D. Kaushik, and S. Patnaik

**Abstract - Magnetoresistance and radio frequency penetration depth techniques are used to study grain connectivity and broadening of superconducting transition. We study these issues in three different superconducting systems e.g. $MgB_2$, $NbSe_2$ and $Bi_2Sr_2Ca_2Cu_3O_{10}$. Some of the samples were irradiated with heavy ions. From the rf response, the bulk pinning force constant is evaluated. From high field transport measurements, H-T phase diagram is compared for the three systems with varying degrees of fluctuation and connectivity.**

## I. Introduction

Thermal fluctuations, anisotropy, and quenched disorder can severely alter the H-T phase diagram of superconductors and thereby limit their potential usage. For example, the copper oxide based high $T_c$ superconductor $Bi_2Sr_2Ca_2Cu_3O_{10}$ (BSCCO) has the upper critical field ($H_{c2}(0)$) in the excess of 150T and transition temperature ($T_c$) ~ 110K but it has not been found suitable for even 1T magnet applications at liquid nitrogen temperature. The high-field magnet industry is still centered around Nb based superconductors that require cooling down to liquid helium temperature. Therefore the 39K superconductor magnesium diboride ($MgB_2$) has renewed the interest for superconducting applications, especially in conjunction with liquid cryogen free close cycle refrigerators [1]. The other issue that is of vital importance is the intergrain connectivity. Eventually this determines the bulk critical current density ($J_c$). By dirtying the sample one can in principle improve $J_c$ and $H_{c2}$. The advantage of $MgB_2$ is that it is a strongly linked superconductor where grain boundaries are more or less transparent to supercurrent flow. But what happens to intergrain connectivity with progressive dirtying? How does the pinning force density compare between the as grown specimen of these three different class of superconductors? In this paper we study and compare these issues in high quality bulk samples of $MgB_2$, BSCCO and $NbSe_2$ ($T_c$ ~7.5K). Our results indicate that $MgB_2$ pinning and thermodynamic properties are intermediate between BSCCO and $NbSe_2$. Moreover the degree of thermal fluctuations and suppression of bulk pinning force constant at elevated temperature are not negligible for the diboride.

Manuscript received August 29, 2006. This work was supported in part by DST-FIST and young scientist project of DST, Govt. of India. Financial support to SDK from CSIR , India is acknowledged. We would like to thank V. Braccini, J. Giencke J. Jiang and I. Naik for providing the samples used in the study.

S. D. Kaushik is with School of Physical Sciences, Jawaharlal Nehru University, New Delhi-110067.

S. Patnaik is with School of Physical Sciences, Jawaharlal Nehru University, New Delhi-110067.

(corresponding author phone: +91-11-26704783; fax: +91-11-26707537; e-mail: spatnaik@ mail.jnu.ac.in).

## II. EXPERIMENT

The clean polycrystalline samples used in this study were grown using different techniques and their properties have been discussed elsewhere [2,3]. The sizes of the samples were 1×3 $mm^2$, 3×4 $mm^2$ and 3×5 $mm^2$ for $MgB_2$, $NbSe_2$ and BSCCO respectively with thickness varying between 200μm to ~1 mm. For the silver sheathed BSCCO tape, silver was removed from the sample by etching with $NH_4OH$ and $H_2O_2$ solution.

Magneto-resistance was measured using linear four-probe resistivity technique. The external magnetic field ranging from 0-6T was applied perpendicular to probe current direction. Data for magnetoresistance were taken during warming cycle and heating rate was kept at 1K/min. All the measurements were carried out using a Cryogenic cryogen-free 1.6K, 8T magnet system in conjunction with a variable temperature insert.

The rf penetration depth study was undertaken by measuring the shift in frequency of a tunnel diode (BD-4) based ultra-stable oscillator with varying temperature or external dc magnetic field $H_{dc}$ [4]. The sample was kept inside an inductor that formed a part of LC circuit of the oscillator (~1.9MHz). The changing magnetization state of the sample is reflected as the shift in frequency. When an external dc field is applied to create the vortices, the superimposed orthogonal rf field can effectively induce tilt motion of the vortices. If the pinning is strong, this tilt motion is not propagated into the bulk and is confined only to the surface. In the other extreme, under negligible pinning, the oscillating vortices can permeate the entire sample. Thus the rf penetration depth becomes a true measure of bulk pinning force constant. At high fields one can also access the flux flow regime using this technique. The effective change in the penetration is given by

$$\delta\lambda(H,T) = \lambda(H,T)-\lambda(0,T) = -G(f(H,T)-f(0,T))$$

here G is a constant geometric factor related to the coil and sample dimension. In the vortex state and in the dilute limit of flux density, the rf penetration depth for periodic pinning potential can be written as [5],

$$\delta\lambda = (\lambda_f^{-2} - (i/2)\lambda_c^{-2})^{-1/2}$$

where $\lambda_c^2 = B\phi_0/\mu_0 k_p$, $\lambda_c$ being the Campbell penetration depth and $\lambda_f^2 = 2B\phi_0/\mu_0\eta\omega$, $\lambda_f$ being the flux flow penetration depth. Here $k_p$ is the bulk pinning force constant or the Labush parameter, $\eta$ is the flux flow viscosity and $\phi_0$ is the flux quantum. Thus the measurement of shift in frequency allows us to determine the rf magnetic penetration depth in the superconducting state and its temperature dependence and in the vortex state it allows us to estimate the pinning force constant and flux flow viscosity. For quantitative estimates geometrical factor G needs to be determined by careful calibration with standard samples.

## III. RESULT AND DISCUSSION

Fig 1 shows the zero field resistivity plotted against the reduced temperature ($T/T_c$). The onset temperature is found to be 7.3K for $NbSe_2$, 39.3K for $MgB_2$ and 109.5K for BSCCO. The behavior in the normal state resistivity reflects the different mechanisms obeyed in different superconductors. For high $T_c$ BSCCO, at optimal doping, a linear resistivity is expected with an extrapolation to zero at T= 0K [6]. The slope $d\rho/dT$ increases as the system is made dirtier with increasing $\rho(T_c)$. The dichalcogenide 2H-$NbSe_2$ is a layered superconductor that exhibits charge density wave formation around 30K with an anisotropy $\gamma$~3. At low temperatures, below CDW transition, the behavior is given by $\rho = \rho(T_c) + aT^3$ which is assigned to strong s-d interband scattering [7]. $MgB_2$ on the other hand is a two band superconductor with normal state properties dominated by 3D isotropic $\pi$ band [8]. It shows metallic behavior prior to the superconducting transition. The normal state resistivity of $MgB_2$ follows Testardi correlation and electron phonon interaction is confirmed [9]. The $\rho(T_c)$, $\Delta\rho$ ($\rho(T=300K) - \rho(T=T_c)$) and residual resistance ratio ($\rho(T=300K) / \rho(T=T_c)$) are estimated to be 0.36 mΩcm, 1.3mΩcm, 4.3 for BSCCO, 3.5 μΩcm, 47μΩcm, 14.4 for $MgB_2$ and 15μΩcm, 180 μΩcm, 12.7 for $NbSe_2$ respectively.

Since we are using polycrystalline samples, the intergrain connectivity determines the bulk critical current density to a large extent. Gandikota et al. have argued that it is $\Delta\rho$ rather than RRR that truly reflects the grain connectivity [10]. In a progressively damaged series of $MgB_2$ thin films we have previously reported increased $\Delta\rho$ and therefore reduced connectivity with irradiation [11]. We found that with increasing $\rho(T_C)$, $\Delta\rho$ increased, a behavior similar to high $T_C$ cuprates. Contrary to reports by Gandikota et al., our experiments indicate that connectivity in $MgB_2$ gets suppressed with dirtying. Using similar analysis we find that for the as grown bulk samples as plotted in figure 1, $MgB_2$ connectivity is intermediate between $NbSe_2$ and BSCCO (after normalization with $\rho(T_C)$).

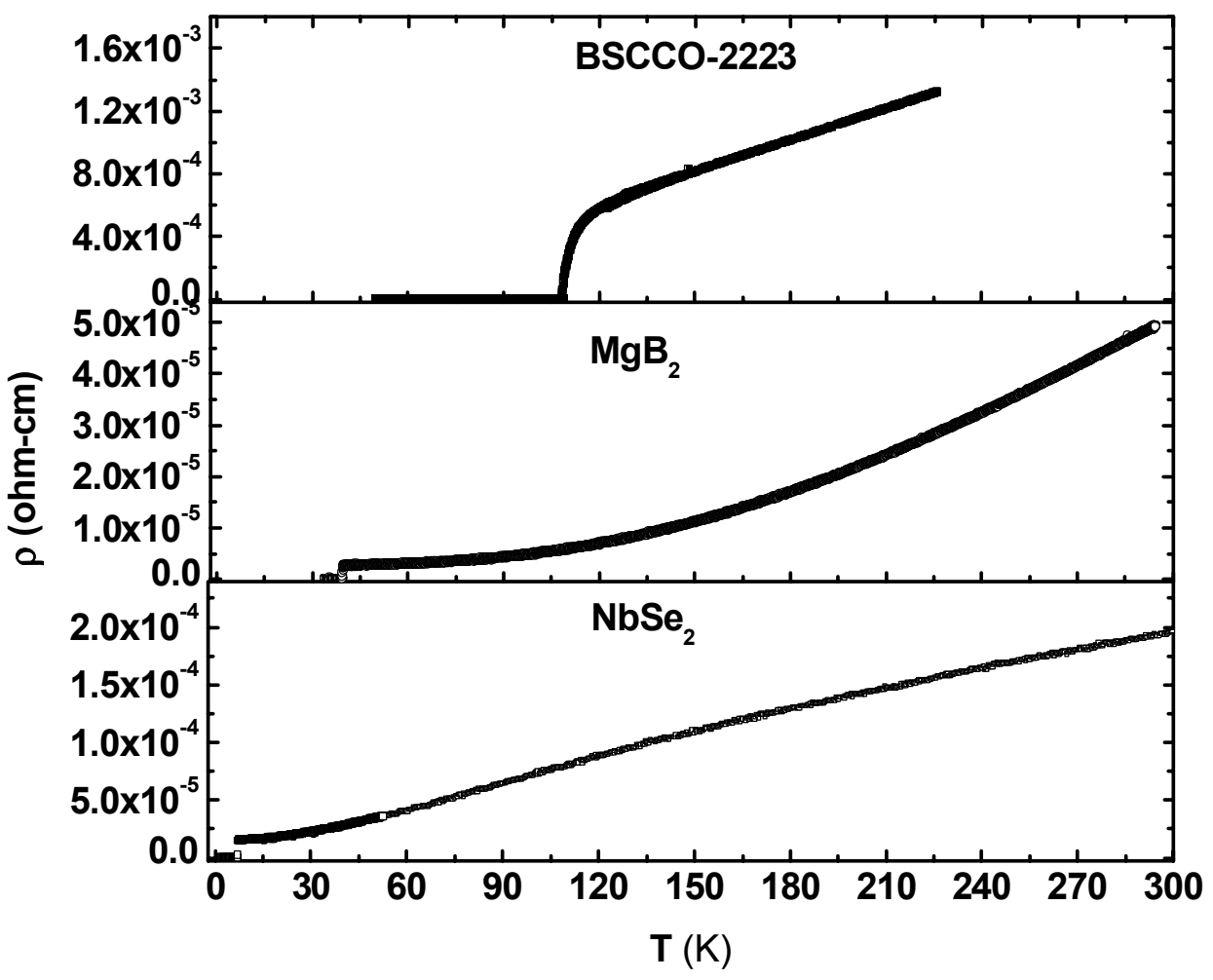

Fig 1: Resistivity behavior from room temperature to low temperature for the three samples. In the normal state BSCCO shows linear behavior, $MgB_2$ shows metallic behavior and $NbSe_2$ shows charge density ordering.

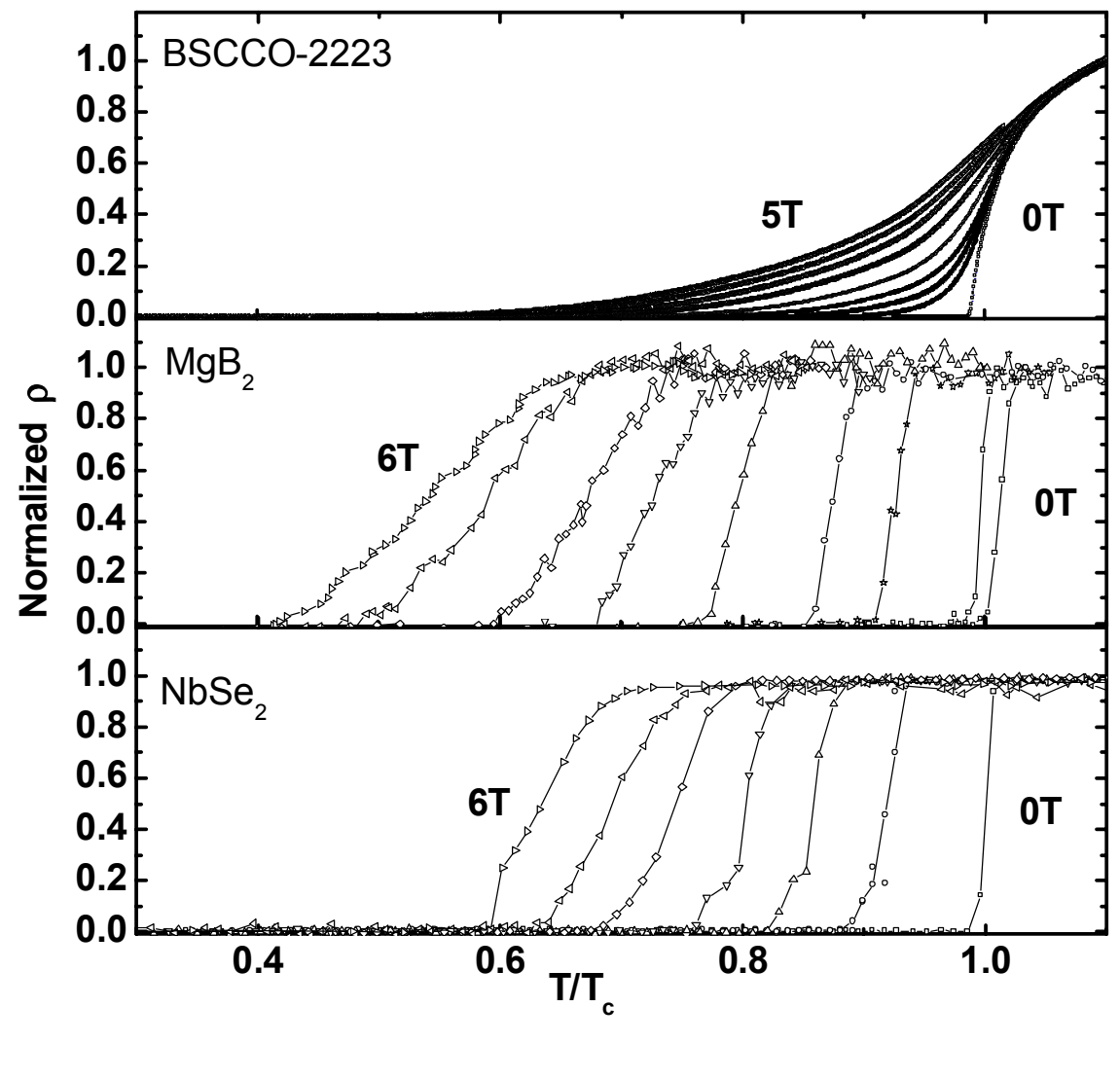

Fig 2: Resistive broadening under magnetic field vs. reduced temperature. Constant magnetic field temperature scans are plotted for BSCCO at 0, 0.1, 0.25, 0.5, 1, 2, 3, 4and 5T, for $MgB_2$ at 0, 0.25, 0.5, 1, 2, 3, 4, 5and 6T, and for $NbSe_2$ at 0, 1, 2, 3, 4, 5, and 6T.

Figure 2 shows the infield transition for the 3 samples. The external field was applied in a direction perpendicular to transport current. The large broadening in resistive transition seen in BSCCO is assigned to severe thermal and quantum fluctuations. Broadening is also seen in $MgB_2$ and $NbSe_2$ although the anisotropy is an order of magnitude lower. In general, broadening can have many sources; e.g. vortex glass phase, surface superconductivity, multigap superconductivity and above all susceptibility to thermal and quantum fluctuations. Unlike BSCCO, the onset $T_C$ shifts with field for the low $T_C$ superconductors. But broadening is not negligible in the boride. For $MgB_2$, above the irreversibility field, using arrhenius analysis we have previously reported that the TAFF activation energy follows a parabolic dependence in B [12].

The resistive broadening seen in Figure 2 is clearly magnetic field dependent. Defining broadening as the $\Delta T$ width between the upper critical and irreversibility points, we plot it as a function of dc magnetic field in Figure 3.

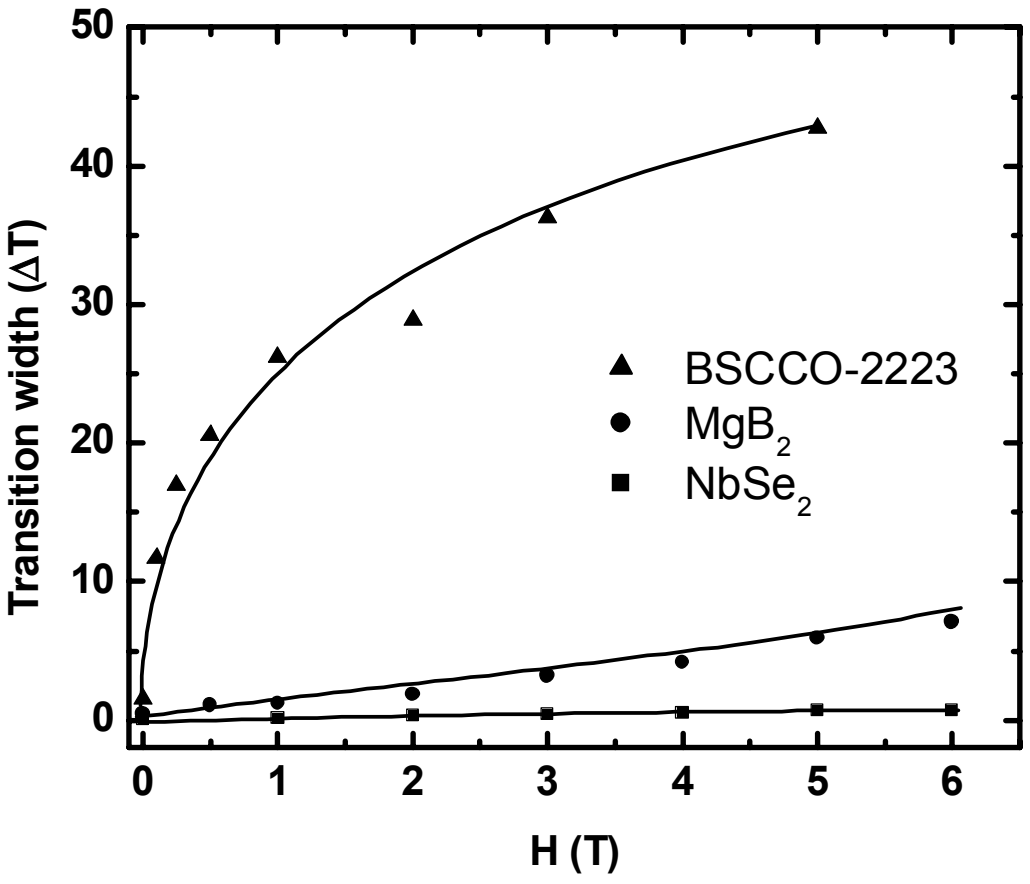

Fig 3: Transition width is plotted against external applied field. Broadening is maximum for BSCCO and minimum for $NbSe_2$.

Clearly BSCCO is most susceptible to broadening. The Ginzberg number $Gi = ( T_c/H_c^2\ \xi_\chi^2\ \xi_{ab}^2)$ gives the quantitative estimate for broadening. For BSCCO it is of the order of $10^{-2}$, while for low $T_c$ superconductors like $NbSe_2$ it could be $10^{-8}$ [13]. $MgB_2$ exhibits intermediate behavior. However the field dependence of broadening cannot be understood entirely as a consequence of difference in coherence length $\xi$.

Plotted in Figure 4 is the upper critical field $H_{C2}$ and irreversibility field $H^*$ as a function of reduced temperature $T/T_C$. The joining lines are guides to eye. The curvature for $MgB_2$ $H_{c2}$ –T is interpreted as dominance of $\sigma$ band [14], where as for BSCCO it marks a 2D to 3D transition. Therefore $dH_{c2}/dT$ near $T_c$ in both the cases do not correlate to $H_{c2}(0)$. The upper critical field in $MgB_2$ is dependent on relative strength of diffusivities between the two bands [14]. The irreversibility field in $MgB_2$ and $NbSe_2$ goes almost linear in temperature where as it shows curvature for BSCCO indicating varying relaxation processes for the vortices.

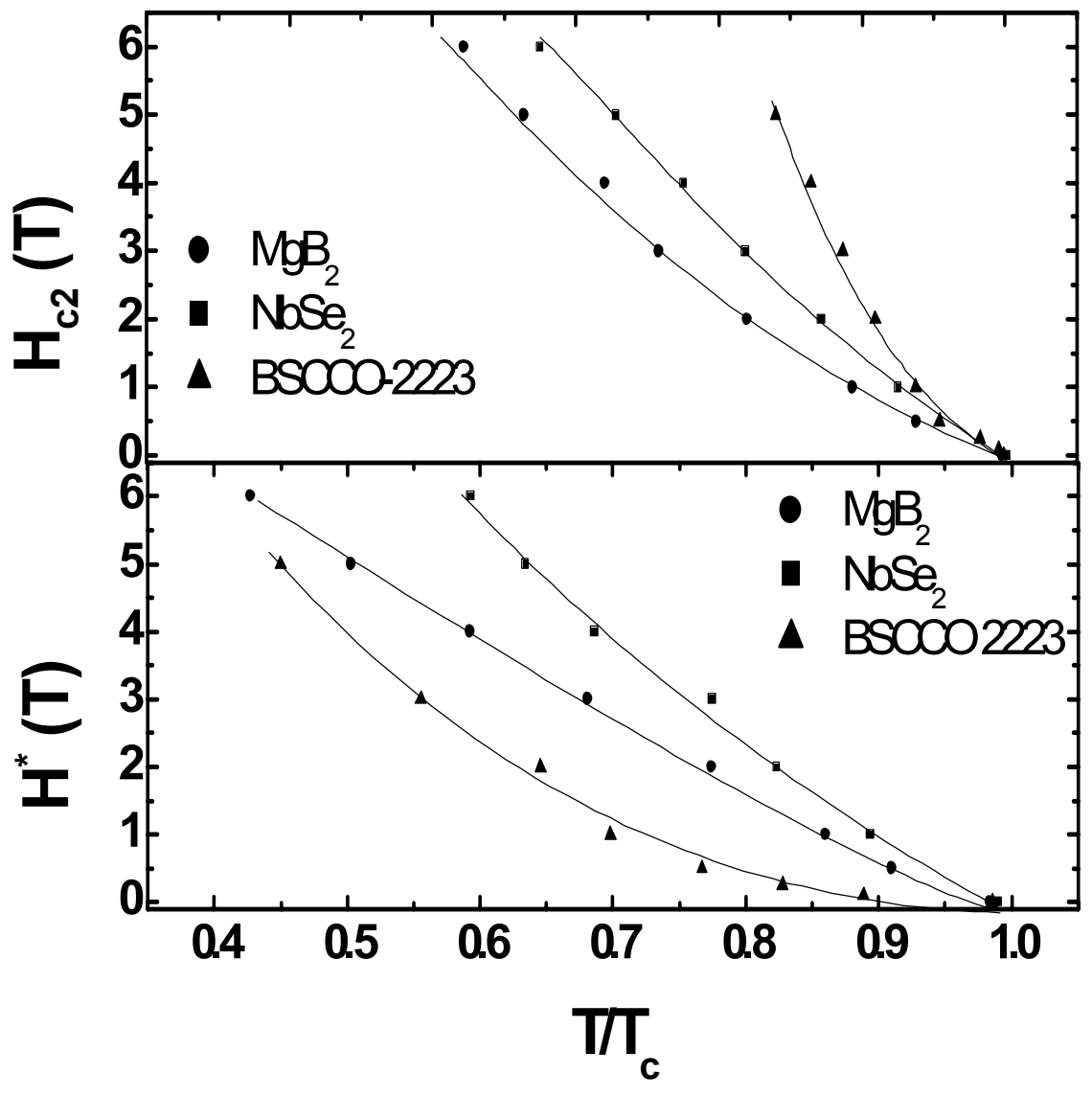

Fig 4: $H_{c2}$ and $H^*$ are plotted for $MgB_2$, $NbSe_2$ and BSCCO against reduced temperature.

Is the broadening seen in all three samples a bulk phenomena? In other words can we rule out surface superconductivity [15]?

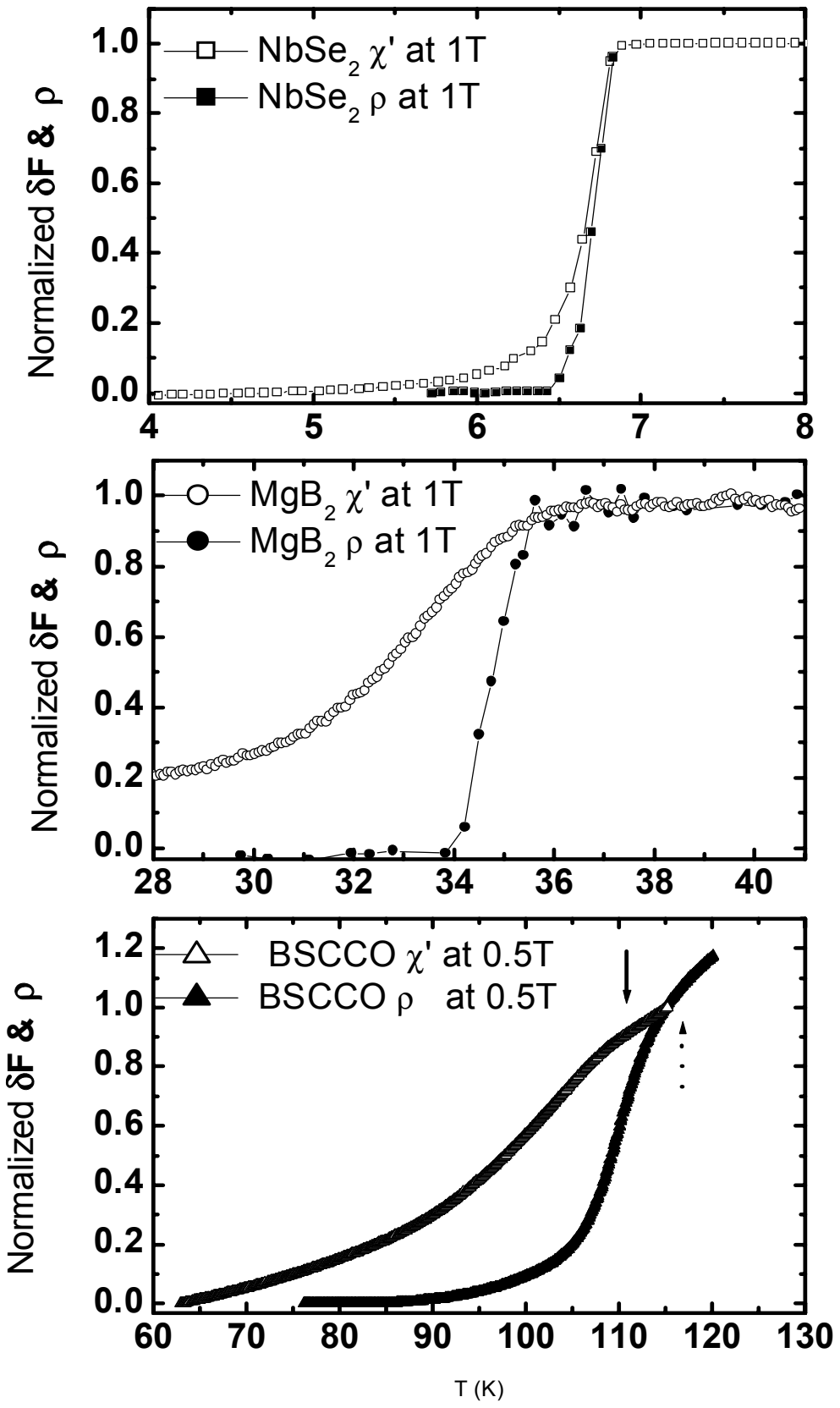

Fig 5: Onset of infield(dc) resistive and diamagnetic transitions occur at the same temperature. The broadening and therefore poor grain connectivity in $MgB_2$ and BSCCO as compared to $NbSe_2$ is evident.

In Figure 5 the magnetization data from normalized δf is plotted as a function of temperature at 1T dc field for $MgB_2$ and $NbSe_2$ and 0.5T for BSCCO. Superimposed are the data from resistivity measurement taken at the same fields. Here $\delta f = -\delta\lambda/G$ is equivalent to $\chi'$ or the inductive part of the susceptibility. Figure 5 clearly shows that onset of broadening is occurring at the same temperature (resistive and magnetization) for all the three superconductors. Moreover the broadening is enhanced in magnetization that reflects poor grain connectivity near the transition temperature. From the broadening we also confirm that $MgB_2$ connectivity is poorer than low $T_c$ $NbSe_2$.

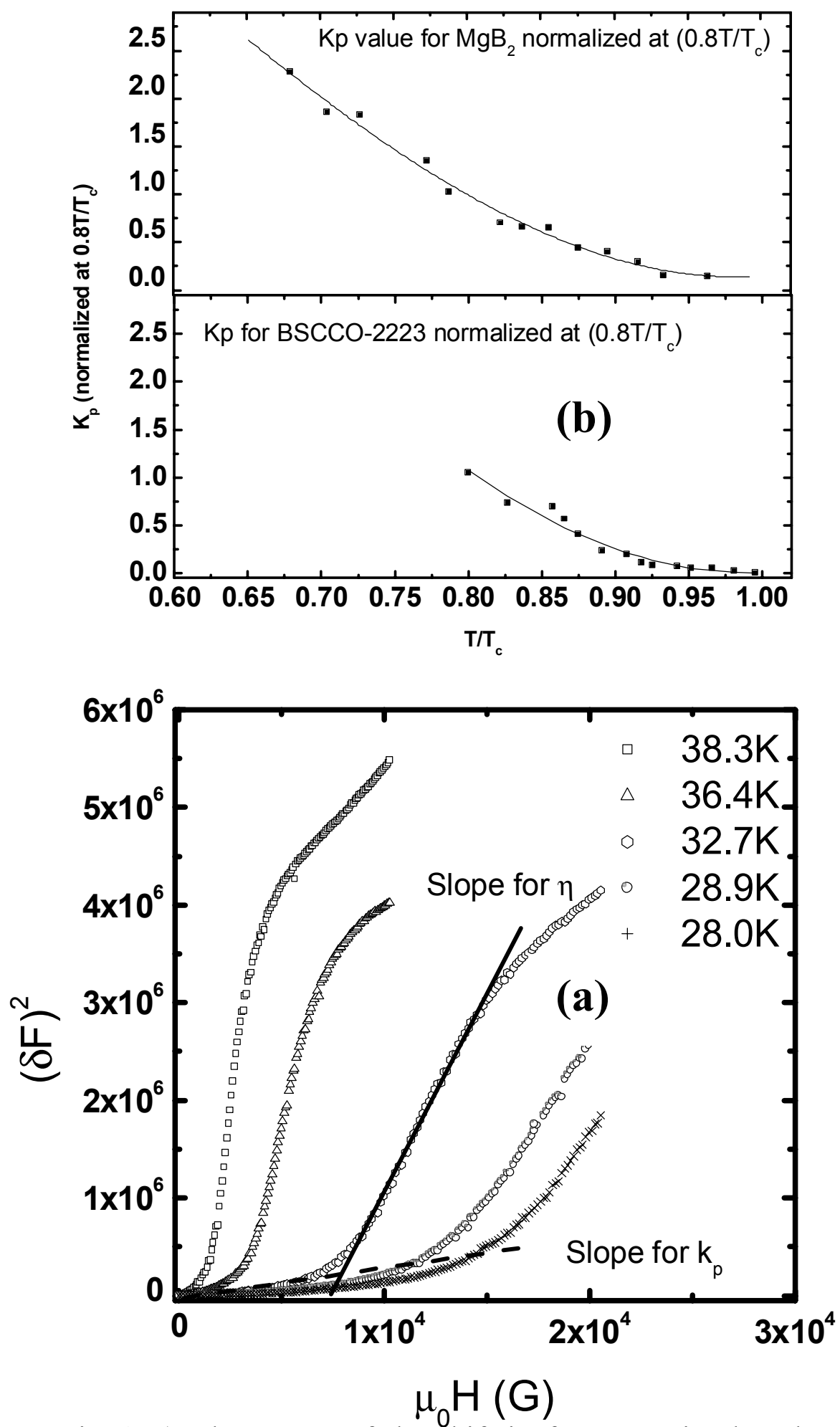

Fig 6: a) The square of the shift in frequency is plotted as a function of magnetic field. The low field slope gives $k_p$. b) Bulk pinning force constant (normalized at $0.8T/T_c$ ) is plotted as a function of reduced temperature. Similar dependence is seen both for $MgB_2$ and BSCCO.

As we discussed in the experimental section, in the vortex state, the pinning force constant can be estimated from rf penetration depth measurements. In Figure 6a we show typical constant temperature field scans for the $MgB_2$ sample. The data has been plotted for $\delta f^2$ vs. H to derive $k_p$ from the slope [16]. Figure 6b shows normalized $k_p$ values as a function of reduced temperature. The plot shows that the temperature dependence of pinning force constant is similar for $MgB_2$ and BSCCO although the magnitude could be vastly different.

In summary, we have studied three superconductors with different thermodynamic and electronic properties and have compared their transport properties. Resistive broadening is seen in all the three systems with varying degree although the Ginzberg number differs by 6 orders of magnitude. As grown $MgB_2$ shows intermediate intergrain connectivity between $NbSe_2$ and BSCCO and shows suppression of connectivity with progressive dirtying. The temperature dependence of bulk pinning force constant shows similar temperature dependence for $MgB_2$ and BSCCO.